\documentclass[preprint,prc,aps,showkeys]{revtex4-1}
\usepackage{amssymb,amsmath}
\usepackage{graphicx,color}

\begin{document} 
\title{Relaxation of particle-hole-type excitation in a Fermi system within the diffusion approximation of kinetic theory for 
the case of constant diffusion and drift coefficients}
\author{Sergiy V. Lukyanov}
\email{lukyanov@kinr.kiev.ua}
\affiliation{\it{Institute for Nuclear Research, 03680 Kyiv, Ukraine}}
%\pacs{21.60.-n, 21.60.Ev, 24.30.Cz}
\date{\today}

\begin{abstract}
The time evolution of the distribution function for a particle-hole excitation in a Fermi system was calculated using the direct numerical solution of 
a nonlinear diffusion equation in momentum space. A phenomenological expression for calculating the relaxation time of such an excitation to its equilibrium value 
has been proposed. It is shown that the relaxation time is dependent on both the excitation energy and the mass number.
\end{abstract}


\maketitle

\section{Introduction}

The study of physical phenomena occurring in finite-size Fermi systems remains an active area of research \cite{KoSh.b.2020}. Among these, the relaxation processes
of collective and single-particle excitations in atomic nuclei are particularly important and intriguing. The relaxation processes in many-particle Fermi systems are conveniently analyzed using quantum kinetic theory methods. Within this framework, the system is described by the kinetic equation for the Wigner distribution function 
in the phase space of coordinates and momenta. The advantage of this approach is that it allows for a relatively straightforward description of average quantities 
such as nucleon density, density flux, pressure, and others in a quantum Fermi system. However, complications arise due to the presence of the collision integral 
on the right-hand side of the kinetic equation. To address these challenges, various simplification methods are employed 
\cite{AbKh.RPP.1959,Be.ZPA.1978,KoLuPlSh.PRC.1998}.

One of these methods is the so-called diffusion approximation \cite{LiPi.bp2.1980}. For instance, in \cite{Wo.PRL.1982}, a schematic model describing 
the equilibrium state in a finite-size Fermi system was considered. The master equation for single-particle states was transformed into a nonlinear 
diffusion equation, taking the Pauli exclusion principle into account. Within this model, an analytical solution was obtained in the simplified case 
of constant transport coefficients. This made it possible to both describe the time evolution of the initial-distribution function for the Fermi system 
and to derive an expression for the equilibrium relaxation time of this excitation: $\tau = 4D/v^2$, where $D$ is the diffusion coefficient 
and $v$ is the drift coefficient. In subsequent studies \cite{BaWo.AP.2019}, the authors presented another, alternative method for solving the nonlinear 
diffusion equation in the approximation of constant kinetic coefficients.

In works \cite{KoLu.UPJ.2014,KoLu.IJMP.2015}, it was demonstrated that the Landau-Vlasov kinetic equation with the nine-dimensional collision integral 
in phase space can be reduced to a diffusion equation for the Wigner distribution function in momentum space. Explicit expressions for the diffusion and drift 
coefficients were obtained. At low temperatures in a Fermi system, the small transferred momentum approximation is used to simplify calculations involving particle 
scattering near the Fermi surface \cite{AbKh.RPP.1959}. However, this assumption alone was insufficient to obtain a physically accurate result. It was necessary 
to make an additional assumption regarding the nature of the interaction between the scattering particles. Previously, the approximation of isotropic nucleon scattering probability was considered adequate \cite{AbKh.RPP.1959,KoLuPlSh.PRC.1998}. However, in this case, the calculation of kinetic coefficients led to divergent integral expressions. This divergence was avoided by imposing a short-range condition on the internucleon potential. Specifically, for a Gaussian potential, it was possible 
to obtain convergent expressions for the kinetic coefficients, calculate their numerical values, and determine their temperature dependencies based solely 
on phenomenological parameters of the internucleon interaction, such as the potential depth and its effective range \cite{KoLu.IJMP.2015}.

The diffusion equation can be used to calculate the time evolution of the distribution function in phase space starting from an initial distribution. 
In the approximation of constant kinetic coefficients, as demonstrated by the authors in \cite{Wo.PRL.1982,BaWo.AP.2019}, this can be done analytically, 
which is one of the rare cases where a nonlinear equation can be solved exactly. In the works \cite{KoLu.UPJ.2014,KoLu.IJMP.2015}, the nonlinear diffusion 
equation was solved through a direct numerical calculation. It is evident that the solution in this case exhibits properties similar to the analytical approach, 
namely, the gradual spreading of the initial-distribution function over time until it reaches the equilibrium Fermi distribution, with a temperature determined 
by the ratio of the kinetic coefficients. However, the advantage of this method is that the direct numerical calculation allows for future solutions in the more 
general case of momentum-dependent kinetic coefficients.

In this paper, Section~\ref{sec-difapp} presents the diffusion approximation for the kinetic equation with the collision integral. In Section~\ref{sec-taueq}, 
a phenomenological method is proposed for calculating the equilibrium relaxation time $\tau_{eq}$ for the initial-distribution function in a spherical Fermi system, 
which models an atomic nucleus in its ground state. Section~\ref{sec-2np} examines the case of nucleon-hole type excitation in an atomic nucleus and uses the proposed 
formula to calculate the dependence of $\tau_{eq}$ on the excitation energy $E_{ex}$ of the nucleon-hole pair and the mass number $A$. 
Conclusions are presented in Section~\ref{concl}.

\section{Diffusion approximation}\label{sec-difapp}

Let us consider the kinetic equation with the collision integral on the right-hand side
\begin{equation}
\frac{\partial f(\mathbf{r},\mathbf{p},t)}{\partial t}+\hat{L}f(\mathbf{r},\mathbf{p},t) =
\mathrm{St}\{f\},
\label{kineq}
\end{equation}
where $f(\mathbf{r},\mathbf{p},t)\equiv f(\mathbf{p})$ is the Wigner distribution function in phase space, $\mathrm{St}\{f\}$ is the collision integral, 
and the operator $\hat{L}$ is given by the expression
\begin{equation}
\hat{L}=\frac{1}{m}\mathbf{p}\cdot\mathbf{\nabla}_\mathbf{r}
-\left(\mathbf{\nabla}_\mathbf{r}U\right)\cdot\mathbf{\nabla}_\mathbf{p},
\label{operator_l}
\end{equation}
where $m$ is the particle mass. In the general case, the single-particle potential $U$ includes both the self-consistent and external fields. 
For the collision integral, we use the expression obtained in the diffusion approximation \cite{KoLu.UPJ.2014,KoLu.IJMP.2015}
\begin{equation}
\mathrm{St}\{f\}= 
- \nabla_{p_\nu} \left[ K_p(\mathbf{p}) f(\mathbf{p}) \left(1-f(\mathbf{p})\right) \frac{p_\nu}{m} + f(\mathbf{p})^2 \nabla_{p_\nu} D_p(\mathbf{p})  \right] 
+ \nabla_{p_\nu}^2\left[f(\mathbf{p}) D_p(\mathbf{p})\right],
\label{stf}
\end{equation}
where the quantities $D_p(\mathbf{p})$ and $K_p(\mathbf{p})$ define the diffusion and drift coefficients in momentum space.

We study a Fermi system that models a spherical atomic nucleus in its ground state. To do this, we consider the following sequence of approximations. 
Let us assume infinite, uniformly distributed nuclear matter in coordinate space. Then, in the kinetic equation (\ref{kineq}), we set 
$f(\mathbf{r},\mathbf{p},t) = f(\mathbf{p},t)$, and thus, $\hat{L}f(\mathbf{r},\mathbf{p},t) = 0$. In this nuclear matter, we isolate a spherical region with radius $R$. 
The distribution function of such a system will exhibit spherical symmetry in momentum space, i.e., $f(\mathbf{p}) = f(p)$. For the expressions $D_p(\mathbf{p})$ and 
$K_p(\mathbf{p})$ in the collision integral (\ref{stf}), we apply the approximation of constant kinetic coefficients 
\cite{Wo.PRL.1982,KoLu.UPJ.2014,KoLu.IJMP.2015,BaWo.AP.2019,Lu.NPAE.2023}, namely: $D_p(\mathbf{p}) = D_{p,0}$ and $K_p(\mathbf{p}) = K_{p,0}$. 
Taking into account the aforementioned approximations, the kinetic equation (\ref{kineq}) transforms into a nonlinear diffusion equation in momentum space.
\begin{equation}
\frac{\partial f(p)}{\partial t}=-\frac{K_{p,0}}{m} \left[ p(1 - 2f(p)) \frac{\partial f(p)}{\partial p} + 3 f(p)(1-f(p))\right] 
+ D_{p,0} \left[ \frac{\partial^2 f(p)}{\partial p^2}+ \frac{2}{p} \frac{\partial f(p)}{\partial p} \right].
\label{eqdif0}
\end{equation}

The diffusion equation (\ref{eqdif0}) must be supplemented with an initial condition. As such, we choose a step distribution function
\begin{equation}
f(p,t=0)=f_{in,0}(p)=\Theta\left(p_F-p\right),
\label{fin0}
\end{equation}
where $p_F$ is the Fermi momentum, which is determined from the condition of particle number conservation
\begin{equation}
\frac{4\pi\mathsf{g}V}{(2\pi\hbar)^{3}}\int_{0}^{p_F} dp\ p^2 =A.
\label{masscons}
\end{equation}
Here, $\mathsf{g} = 4$ is the degeneracy factor for nucleons, and $V = 4\pi R^3 / 3$ is the volume of the nucleus, where its radius $R$ is related to the mass number $A$ 
by the relation $R = r_0 A^{1/3}$. In the subsequent calculations, we use the value for the coefficient $r_0 = 1.2$ fm. After integrating over momentum and 
substituting the expression for the nucleus volume, we obtain the expression for the Fermi momentum
\begin{equation}
p_F=\left(\frac{9\pi}{8}\right)^{1/3}\frac{\hbar}{r_0},
\label{pf}
\end{equation}
where $\hbar$ is the Planck constant. Substituting the numerical values of $\hbar$ and the coefficient $r_0$ into equation (\ref{pf}), 
we obtain the numerical value of the Fermi momentum: $p_F\approx 8.4\times 10^{-22}$ $\text{MeV}\cdot\text{fm}^{-1}\cdot $s. This value of the Fermi momentum $p_F$ 
corresponds to the Fermi energy: $E_F=p_F^2/2m\approx 33.7$ MeV, where $m$ is the nucleon mass.

When performing numerical calculations, we use the values of the kinetic coefficients obtained in our work \cite{Lu.NPAE.2023}. In the approximation of constant 
coefficients, their values are calculated at momenta equal to the Fermi momentum, such that $D_{p,0} = D_p(p_F)$ and $K_{p,0} = 3K_p(p_F)$. The value of the diffusion 
coefficient $D_{p,0} \approx 3.4 \times 10^{-22}$ MeV$^{2} \cdot$fm$^{-2} \cdot$ s describes the experimental data well and is consistent with the estimates 
of other authors \cite{Wo.PRL.1982,BaWo.AP.2019,Lu.conf.KINR2023}. For the numerical value of the drift coefficient, we use $K_{p,0} \approx -8.3 \times 10^{-23}$ 
MeV$\cdot$fm$^{-2}\cdot$ s, which satisfies the relation between the equilibrium temperature $T_{eq}$ and the kinetic coefficients (see relation (\ref{teq})).

\begin{figure}
\begin{center}
\includegraphics[scale=0.6,clip]{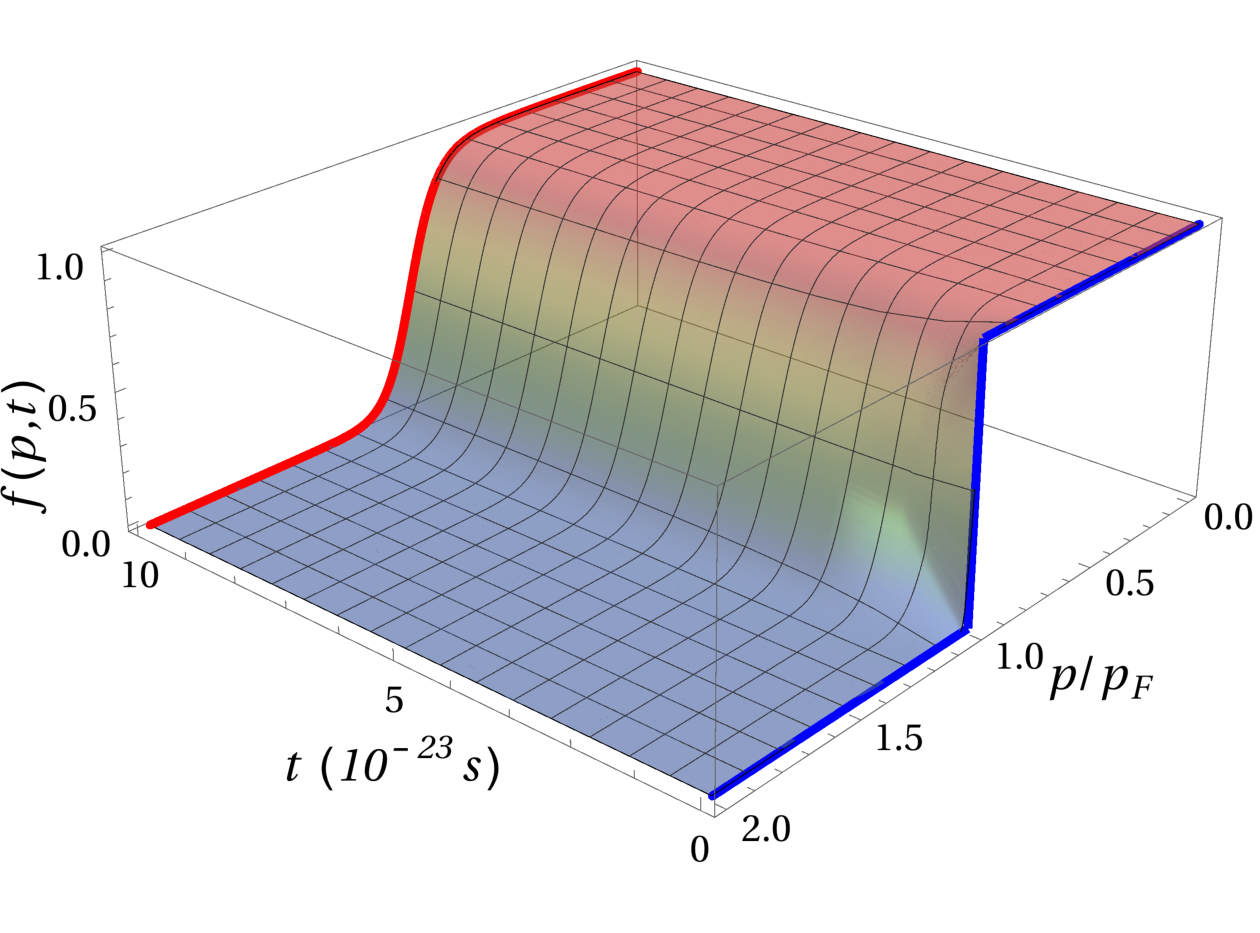}
\vspace{-0.6cm}
\caption{Time evolution of the distribution function $f(p,t)$ obtained using the diffusion equation (\ref{eqdif0}). 
The blue line represents the initial-distribution function (\ref{fin0}), while the red curve represents the equilibrium Fermi-distribution function (\ref{feq}).}
\label{fig1}
\end{center}
\end{figure}

In \figurename~\ref{fig1}, the time dependence of the distribution function $f(p,t)$ calculated using the diffusion equation (\ref{eqdif0}) is shown. 
The momentum dependence is presented in units of the Fermi momentum $p/p_F$, so the Fermi surface corresponds to a unit value of the relative momentum. 
As seen from the figure, the obtained distribution function $f(p,t)$ gradually broadens over time and evolves toward the equilibrium Fermi distribution.
\begin{equation}
f_{eq}(p)=\left(1+\exp\frac{p^2/2m-E_{F,eq}}{T_{eq}}\right)^{-1},
\label{feq}
\end{equation}
where $E_{F,eq}$ is the Fermi energy, which differs from the Fermi energy $E_{F}$ for the step distribution and is determined from the condition
\begin{equation}
\frac{4\pi\mathsf{g}V}{(2\pi\hbar)^{3}}\int_{0}^{\infty} dp\ p^2 f_{eq}(p) =A.
\label{massconsfeq}
\end{equation}
Here, $T_{eq}$ is the equilibrium temperature, which, according to \cite{Lu.IJMP.2021,Wo.PRL.1982,BaWo.AP.2019}, is determined by the ratio of kinetic coefficients
\begin{equation}
T_{eq}= -\frac{D_{p,0}}{K_{p,0}}.
\label{teq}
\end{equation}
At the given values of the kinetic coefficients, the equilibrium temperature is $T_{eq}\approx 4$ MeV. For this equilibrium temperature, from equation (\ref{massconsfeq}), 
we numerically find the Fermi energy for the equilibrium distribution (\ref{feq}): $E_{F,eq}\approx 33.3$ MeV. As we can see, the obtained value $E_{F,eq} < E_F$, 
which is due to the diffuseness of the equilibrium distribution function.

\section{Relaxation time}\label{sec-taueq}

Let us consider the deviation of the distribution function from its equilibrium value
\begin{equation} 
\delta f(p,t)=f(p,t)-f_{eq}(p). 
\label{df} 
\end{equation} 
It is clear that at the initial moment, $t=0$, the distribution function is in its initial state
\begin{equation} 
f(p,t=0)=f_{in}(p). 
\end{equation} 
We then denote the initial deviation from the equilibrium distribution as
\begin{equation}
\delta f(p,t=0)\equiv\delta f_{in}(p)=f_{in}(p)-f_{eq}(p).
\end{equation}

Let us introduce the root mean square deviation
\begin{equation}
\Delta (t) = \sqrt{\int d\mathbf{p}\ \left[ \delta f(p,t) \right]^2}. 
\label{deltat}
\end{equation}
Its initial value will be
\begin{equation}
\Delta (t=0)\equiv\Delta_{0}= \sqrt{\int d\mathbf{p}\ \left[ \delta f_{in}(p) \right]^2}.
\label{delta0}
\end{equation}

In the future, it will be more convenient to consider not the function $\Delta(t)$ itself, but the quantity normalized to its initial value, $\Delta(t)/\Delta_0$. 
Clearly, at time $t=0$, the ratio $\Delta(t)/\Delta_0$ equals one, indicating that the deviation of the distribution function from its equilibrium value is at its 
maximum.

\begin{figure}
\begin{center}
\includegraphics[scale=0.5,clip]{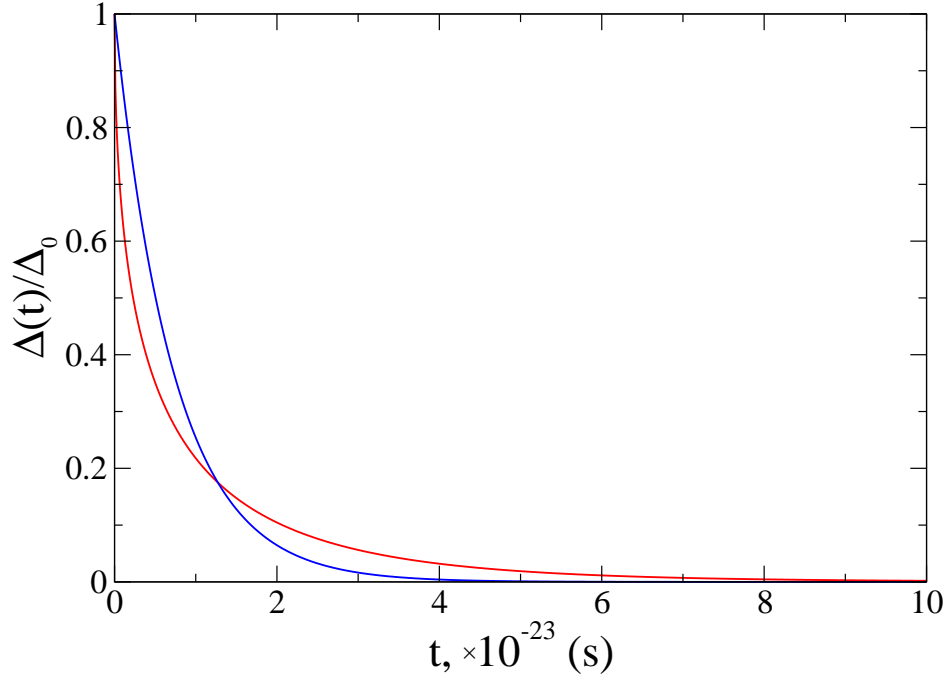}
\caption{The dependence of the ratio $\Delta(t)/\Delta_0$ on time $t$. The red curve represents the numerical calculation, while the blue curve corresponds 
to the exponential dependence (\ref{delta4}) with a relaxation time $\tau_{eq} \approx 7.3 \times 10^{-24}$ s.}
\label{fig2}
\end{center}
\end{figure}

In \figurename~\ref{fig2}, the calculated dependence of the ratio $\Delta(t)/\Delta_0$ on time (red curve), according to expressions (\ref{df})-(\ref{delta0}) 
is shown for the previously obtained time evolution of the distribution function, depicted in \figurename~\ref{fig1}.
As can be seen from the figure, over time, $\Delta(t)/\Delta_0$ decreases and asymptotically approaches zero. This dependence is nonexponential in nature.

Let us assume that the relaxation of the deviation of the distribution function from its equilibrium value is described by an exponential dependence 
\begin{equation} 
\delta f(p,t)=\delta f_{in}(p) \exp\left(-t/\tau_{eq}\right), 
\label{dftaueq} 
\end{equation} 
where $\tau_{eq}$ is the relaxation time. Substituting (\ref{dftaueq}) into expression (\ref{deltat}), we obtain 
\begin{equation} 
\Delta (t)=\Delta_0 \exp\left(-t/\tau_{eq}\right). 
\label{delta4} 
\end{equation} 
Integrating both sides of expression (\ref{delta4}) with respect to time yields 
\begin{equation} \
\int_0^\infty\Delta (t) dt=\Delta_0 \int_0^\infty\exp\left(-t/\tau_{eq}\right)dt. 
\label{delta4-intt} 
\end{equation} 
By performing the integration on the right-hand side of (\ref{delta4-intt}), we obtain 
\begin{equation} 
\tau_{eq}= \int_0^\infty \frac{\Delta (t)}{\Delta_0} dt. 
\label{taueq} 
\end{equation} 
Substituting the exponential dependence (\ref{delta4}) into this expression gives a trivial identity. However, in the case of a nonexponential time dependence 
of $\Delta(t)$, we obtain a value of $\tau_{eq}$ that characterizes the relaxation time in this scenario.

Thus, by substituting expressions (\ref{deltat}) and (\ref{delta0}) into (\ref{taueq}), one can obtain the relaxation time for an arbitrary time dependence 
of the root-mean-square deviation of the distribution function from its equilibrium value. In this case, the area under the curve $\Delta(t)/\Delta_0$ will 
be equal to the area under the exponential dependence $\exp\left(-t/\tau_{eq}\right)$ and will correspond to the relaxation time $\tau_{eq}$ calculated using 
formula (\ref{taueq}). In \figurename~\ref{fig2}, the exponential dependence $\exp\left(-t/\tau_{eq}\right)$ with a relaxation time 
$\tau_{eq} \approx 7.3 \times 10^{-24}$ s, obtained using (\ref{taueq}) and the expressions (\ref{deltat}) and (\ref{delta0}) for the evolution of the distribution 
function depicted in \figurename~\ref{fig1}, is shown in blue. As can be seen from the figure, the areas enclosed by the blue and red curves are equal.

\section{Particle-hole-type excitation}\label{sec-2np}

Let us consider the case of a particle-hole-type excitation, which is described by an initial-distribution function of the form 
\begin{eqnarray}
f_{\mathrm{in}}(p) &=&\left[ 1-\Theta (p-p_{1}^{\prime }) +\Theta(p-p_{2}^{\prime })\right] \left[ 1-\Theta (p_F-p)\right]  \nonumber \\
& &+\left[ 1-\Theta (p-p_{2})\right] \Theta (p-p_{1})\Theta (p_F-p).
\label{fin1p1h}
\end{eqnarray} 
The distribution (\ref{fin1p1h}) indicates that the particle is localized beyond the Fermi surface with momentum $p_1>p_F$, while the hole is accordingly located 
below the surface with $p'_1<p_F$. The width of the interval in momentum space, describing the particle, $\Delta p=p_2-p_1$, is determined from the following condition
\begin{equation}
\frac{4\pi\mathsf{g}V}{(2\pi\hbar)^{3}}\int_{p_F}^{\infty}dp\ p^{2} f_{in}(p)=1,  
\label{fin1p}
\end{equation} 
and the width of the interval describing the hole, $\Delta p'=p'_2-p'_1$, is determined accordingly from the condition
\begin{equation}
\frac{4\pi\mathsf{g}V}{(2\pi\hbar)^{3}}\int_{0}^{p_F}dp\ p^{2} f_{in}(p)= A-1.
\label{fin1h}
\end{equation}
Note that these two conditions automatically satisfy the conservation of the total particle number. It is also worth mentioning that the Fermi momentum in this case 
is not equal to the Fermi momentum for the step distribution (\ref{fin0}). However, their difference is very small, so we will neglect it.

\begin{figure}
\begin{center}
\includegraphics[scale=0.6,clip]{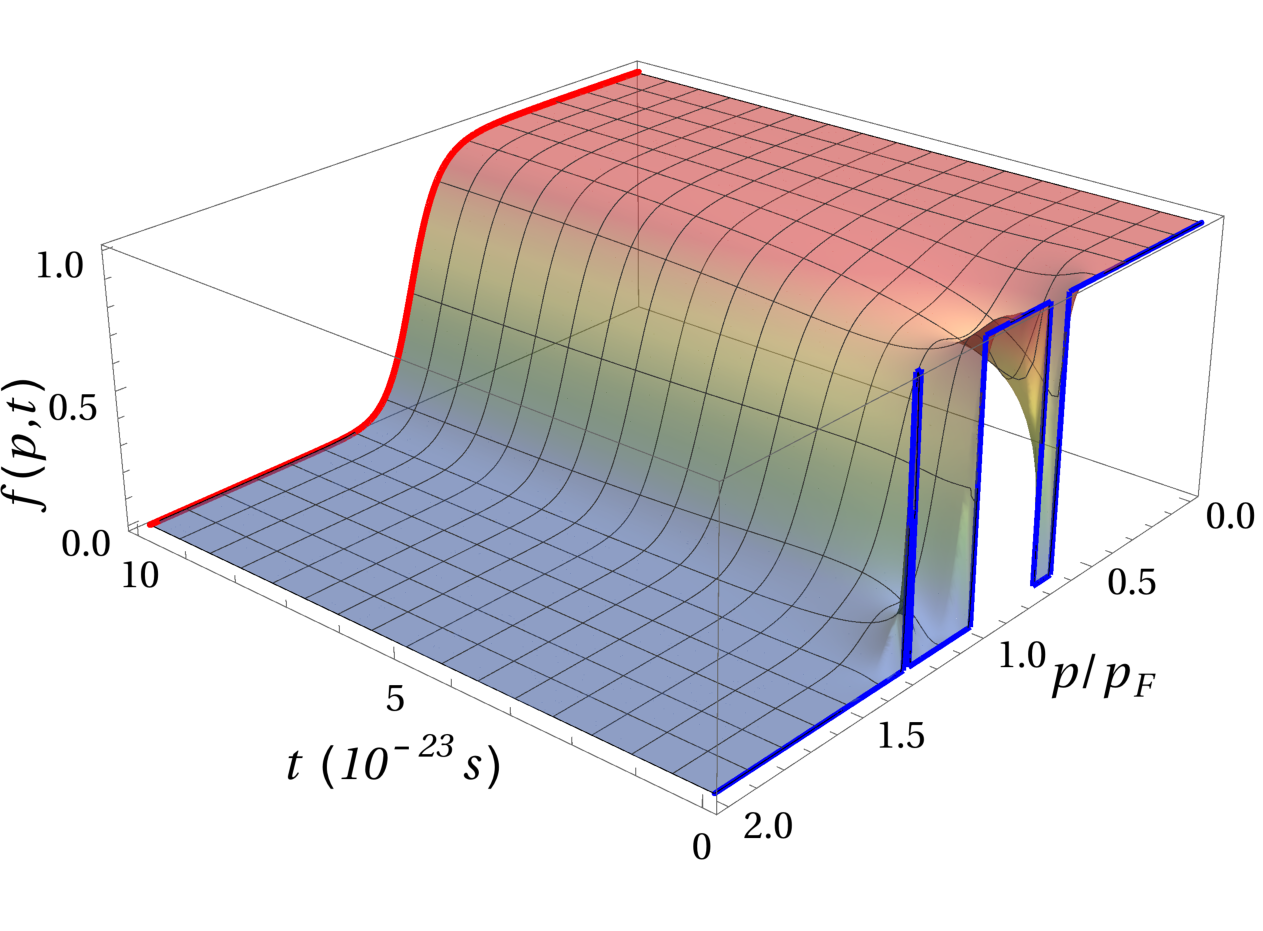}
\vspace{-0.6cm}
\caption{The illustration of the time evolution of the distribution function $f(p,t)$ for the initial distribution describing the particle-hole-type excitation 
(\ref{fin1p1h}) is shown by the blue line. The red line represents the equilibrium Fermi-distribution function (\ref{feq}) at temperature of $T_{eq}=4$ MeV.}
\label{fig3}
\end{center}
\end{figure}

The time evolution of the distribution function for the initial particle-hole-type distribution (\ref{fin1p1h}) is illustrated in \figurename~\ref{fig3}.
As can be seen from the figure, similar to the previous case, the initial-distribution function becomes blurred and eventually approaches the equilibrium 
distribution (\ref{feq}). According to the calculations in Appendix A, the widths of the intervals in momentum space, which describe the particle and the hole, 
are inversely proportional to the mass number and the square of the momentum: $\Delta p \approx p_F^3/(3 A p^2_1)$ and $\Delta p' \approx p'^3_F/(3 A p'^2_1)$.
Therefore, as shown in the figure, the relative width of the dip associated with the hole, $\Delta p'/p_F$, is greater than the relative width of the peak 
associated with the particle, $\Delta p/p_F$, since the widths in both cases are inversely proportional to the squares of the momenta of the hole and the particle.
It should be noted that, due to the smallness of the relative widths $\Delta p/p_F$ and $\Delta p'/p_F$ for heavy and medium nuclei, $A=10$ was chosen for clarity 
in calculating the evolution of the distribution function. Additionally, to separate the peak and dip from the Fermi surface background, arbitrary energy values 
for the particle and hole, equidistant from the Fermi surface, were chosen. 

The excitation energy of the particle-hole pair, $E_{ex}$, is the difference between the energies of the nucleon, $E_1$, and the hole, $E'_1$. In this study, 
the energy of the hole was chosen arbitrarily, but with consideration of its small deviation from the Fermi energy. Specifically, we chose $E'_1 = E_F - 2$ MeV 
arbitrarily.

The average binding energy per nucleon in atomic nuclei is approximately $B/A \approx 8$ MeV. Therefore, it makes sense to consider excitation energies, 
$E_{ex}$, where the excited nucleon remains bound within the nucleus, i.e., $E_{ex} - 2 \leq B/A$. Given this constraint, we will focus on the excitation 
of the nucleon-hole pair within the range $2 \leq E_{ex} \leq 10$ MeV.

\begin{figure}
\begin{center}
\includegraphics[scale=0.5,clip]{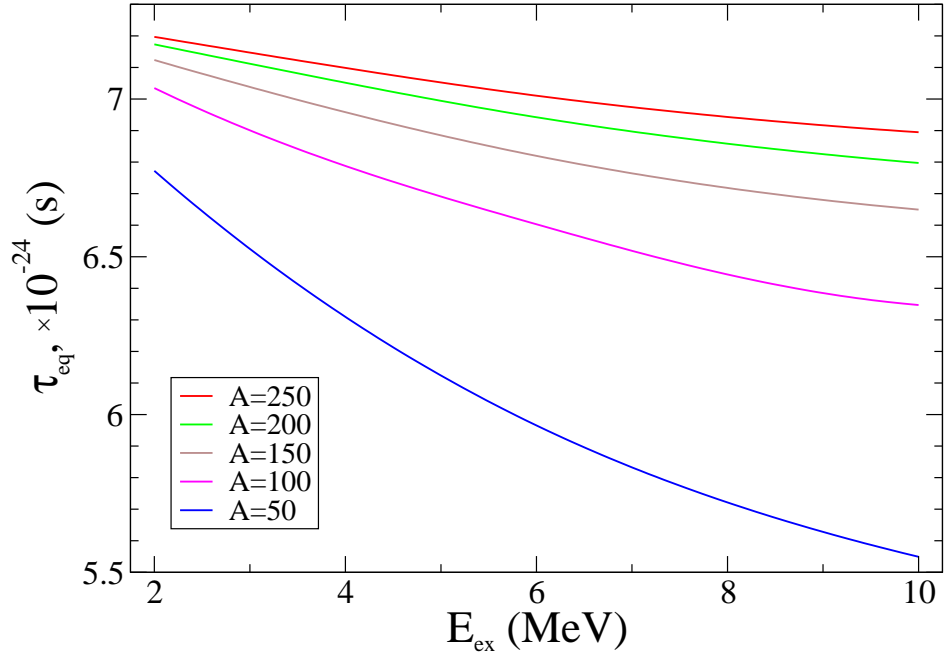}
\caption{The relaxation time $\tau_{eq}$ as a function of the nucleon excitation energy $E_{ex}$ for nuclei with different mass numbers. 
The nucleon is excited from the $E_F-2$ MeV level.}
\label{fig4}
\end{center}
\end{figure}

\figurename~\ref{fig4} shows the dependence of the relaxation time, $\tau_{eq}$, calculated according to expression (\ref{taueq}), 
for the particle-hole-type distribution as a function of the excitation energy, $E_{ex}$, for nuclei with different mass numbers.
As seen in the figure, for all nuclei, the relaxation time, $\tau_{eq}$, decreases non-linearly with increasing excitation energy, $E_{ex}$.

\begin{figure}
\begin{center}
\includegraphics[scale=0.5,clip]{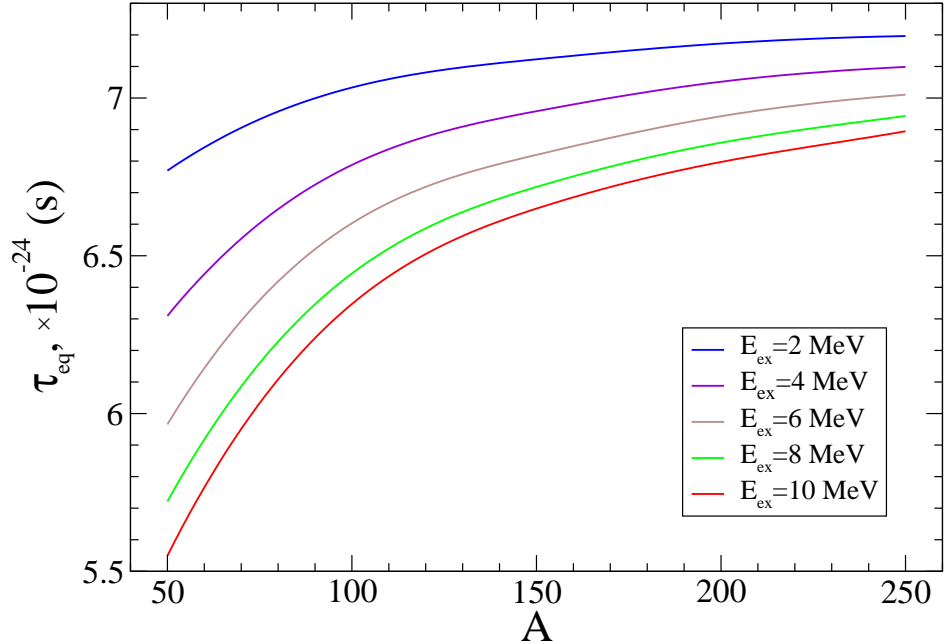}
\caption{The relaxation time $\tau_{eq}$ as a function of the mass number $A$ at different values of the particle-hole excitation energy $E_{ex}$.
A nucleon is excited from the $E_F-2$ MeV level.}
\label{fig5}
\end{center}
\end{figure}

It should also be noted that at all excitation energies, the value of $\tau_{eq}$ increases with the mass number, $A$. 
This dependence is illustrated in \figurename~\ref{fig5}, 
where the relaxation time, $\tau_{eq}$, is plotted as a function of the mass number for different values of the particle-hole excitation energy, $E_{ex}$. 
As seen in the figure, the relaxation time increases with the mass number at all excitation energies.

Finally, it is worth noting that the numerical calculations performed are technically resource-intensive, requiring substantial computational power 
to achieve satisfactory accuracy in the obtained results.

\section{Conclusions}\label{concl}

In this work, within the framework of the diffusion approximation in kinetic theory, a nonlinear diffusion equation for the distribution function was derived for the case of constant diffusion and drift coefficients. Through direct numerical calculations, the time evolution of the distribution function was obtained for both an initial step-like distribution and a case of particle-hole excitation in a Fermi system, modeling the ground state of spherically symmetric atomic nuclei. It was demonstrated that in both cases, the initial distribution gradually blurs over time, evolving toward the equilibrium Fermi distribution, whose temperature, as is known, is determined by the ratio of the diffusion and drift coefficients.

We proposed a phenomenological method for calculating the relaxation time of an arbitrary initial excitation to its equilibrium value. The relaxation time is determined by the area under the equivalent exponential time dependence, normalized to the initial value of the root-mean-square deviation of the distribution function. Using this method, the relaxation time for the step-like initial distribution was calculated as $\tau_{eq} \approx 7.3 \times 10^{-24}$ s.

We also calculated the dependence of the relaxation time on the particle-hole excitation energy in a spherically symmetric Fermi system, where the arbitrarily chosen hole energy is close to the Fermi energy (in our case, the difference is 2 MeV). The calculations showed that as the excitation energy increases, the relaxation time decreases in a nonlinear manner. This trend holds for all values of the mass number. Additionally, it was demonstrated that as the mass number increases, the relaxation time also increases across different values of the particle-hole excitation energy.

Theoretical studies revealed several issues that require further investigation. In particular, the relaxation time, $\tau_{eq}$, obtained using this method was 
approximately an order of magnitude smaller than expected. It can be hypothesized that the reason for this discrepancy may lie in the method used to determine 
the relaxation time. Another possible explanation is that the momentum-integrated quantity $\Delta(t)/\Delta_0$ likely reaches equilibrium much faster 
than the distribution function at the Fermi momentum itself does. This occurs because, in a large fraction of momentum space, the initial distribution is 
already equal to the equilibrium distribution due to the Pauli principle. Additionally, further clarification is needed regarding the energy imbalance for 
the initial particle-hole distribution, as it leads to the equilibrium temperature, $T_{eq}$, being independent of the excitation energy, $E_{ex}$.
\bigskip

\acknowledgments

The author expresses gratitude to the Armed Forces of Ukraine for ensuring safety during the conduct of this research. 
The author sincerely thanks A.I. Sanzhur for useful and creative discussions.

\bigskip

\appendix\section{Conditions for widths in momentum space for a particle and a hole}

Let us examine the conditions (\ref{fin1p}) and (\ref{fin1h}) in more detail. We denote the corresponding integrals as $I_1$ and $I_2$. 
Then, the expressions (\ref{fin1p}) and (\ref{fin1h}) will take the form of
\begin{eqnarray}
\frac{4\pi\mathsf{g}V}{(2\pi\hbar)^{3}}\ I_1 & = & A-1, 
\label{umova1} \\
\frac{4\pi\mathsf{g}V}{(2\pi\hbar)^{3}}\ I_2 & = & 1.   
\label{umova2}
\end{eqnarray}

For the distribution function (\ref{fin1p1h}), the integral $I_1$ can be expressed as follows:
\begin{equation}
I_1=\int_{0}^{p_F}dp\ p^{2} f_{in}(p) = \int_{0}^{p'_1} dp\ p^{2} + \int_{p'_2}^{p_F} dp\ p^{2} =\frac{1}{3}\left(p'^3_1+p^3_F-p'^3_2 \right).
\label{i1}
\end{equation}
Let us express the momentum $p_2'$ through the initial momentum $p_1'$ and the width of the interval of the hole in momentum space $\Delta p'$, 
then $p'_2=p'_1+\Delta p'$. Thus
\begin{equation}
p_2'^3=\left(p'_1+\Delta p'\right)^3=p'^3_1+3p'^2_1\Delta p'+3p'_1\Delta p'^2+\Delta p'^3.
\label{ps23}
\end{equation}
After substituting this expression into (\ref{i1}), we will have
\begin{equation}
I_1=\frac{p^3_F}{3}\left(1-3x'^2_1 \Delta x' -3 x'_1 \Delta x'^2-\Delta x'^3 \right),
\label{integralI12}
\end{equation}
where the notation is introduced: $x'_1=p'_1/p_F$,  $\Delta x'=\Delta p'/p_F$.

Similarly, we will write the integral $I_2$
\begin{equation}
I_2=\int_{p_F}^\infty dp\ p^{2} f_{in}(p) = \int_{p_1}^{p_2} dp\ p^{2} =\frac{1}{3}\left(p^3_2-p^3_1 \right).
\label{i2}
\end{equation}
Let us express the momentum $p_2$ through the initial momentum $p_1$ and the width of the particle interval in momentum space $\Delta p$, then
$p_2=p_1+\Delta p$. Thus
\begin{equation}
p^3_2=\left(p_1+\Delta p\right)^3=p^3_1+3p^2_1\Delta p+3p_1\Delta p^2+\Delta p^3.
\end{equation}
After substituting this expression into (\ref{i2}), we will have
\begin{equation}
I_2=\frac{p^3_F}{3}\left(3x^2_1 \Delta x +3 x_1 \Delta x^2+\Delta x^3 \right).
\label{integralI22}
\end{equation}
where the notation is introduced: $x_1=p_1/p_F$, $\Delta x=\Delta p/p_F$.

After substituting the obtained expressions for the integrals (\ref{integralI12}) and (\ref{integralI22}) into expressions (\ref{umova1}) and (\ref{umova2}), 
we obtain the following conditions
\begin{equation}
\frac{4\pi\mathsf{g}V}{(2\pi\hbar)^{3}} \frac{p^3_F}{3}\left(1-3x'^2_1 \Delta x' -3 x'_1 \Delta x'^2-\Delta x'^3 \right) = A-1,
\label{condition12}
\end{equation}
\begin{equation}
\frac{4\pi\mathsf{g}V}{(2\pi\hbar)^{3}}\ \frac{p^3_F}{3}\left(3x^2_1 \Delta x +3 x_1 \Delta x^2+\Delta x^3 \right)=1.
\label{condition22}
\end{equation}

Since $\Delta x' \ll 1$ and $\Delta x \ll 1$, it makes sense in expressions (\ref{condition12}) and (\ref{condition22}) to consider only the contribution 
from the linear terms
\begin{equation}
\frac{4\pi\mathsf{g}V}{(2\pi\hbar)^{3}} \frac{p^3_F}{3}\left(1-3x'^2_1 \Delta x' \right) \approx A-1,
\label{condition12linear}
\end{equation}
\begin{equation}
\frac{4\pi\mathsf{g}V}{(2\pi\hbar)^{3}}\ \frac{p^3_F}{3}\ 3x^2_1 \Delta x \approx 1.
\label{condition22linear}
\end{equation}

We will consider that with sufficient accuracy, the Fermi momentum $p_F$ can be determined using expression (\ref{masscons}) for the step distribution
\begin{equation}
\frac{4\pi\mathsf{g}V}{(2\pi\hbar)^{3}} \frac{p_F}{3} \approx A.
\label{masscons1}
\end{equation}

Therefore, taking into account the law of conservation of particle number (\ref{masscons1}), the conditions (\ref{condition12linear}) 
and (\ref{condition22linear}) will be written in their final form as
\begin{equation}
\Delta x' \approx 1/(3 A x'^2_1),
\label{umova13}
\end{equation}
\begin{equation}
\Delta x \approx 1/(3 A x^2_1).  
\label{umova23}
\end{equation}

\bibliographystyle{apsrev4-1}
\bibliography{tauex_2}

\end{document}